\documentclass{article}

\usepackage{microtype,pifont}
\usepackage{graphicx}
\usepackage{subfigure}
\usepackage{booktabs} % for professional tables

\usepackage{hyperref}

\usepackage[accepted]{icml2023}

\usepackage{amsmath}
\usepackage{amssymb}
\usepackage{mathtools}
\usepackage{amsthm}

\usepackage[capitalize,noabbrev]{cleveref}

\theoremstyle{plain}

\theoremstyle{definition}

\theoremstyle{remark}

\newcommand{\cmmc}{\textsc{\small 21CMMC}}
\newcommand{\cmfst}{\textsc{\small 21cmFAST}}
\newcommand{\ttan}{\textsc{\small THESAN}}

\usepackage[textsize=tiny]{todonotes}

\icmltitlerunning{21cm Few-shot GAN}

\begin{document}

\twocolumn[
\icmltitle{Multi-fidelity Emulator for Cosmological Large Scale 21 cm Lightcone Images:\\ a Few-shot Transfer Learning Approach with GAN}




\begin{icmlauthorlist}
\icmlauthor{Kangning Diao}{to}
\icmlauthor{Yi Mao}{to}
\end{icmlauthorlist}

\icmlaffiliation{to}{Department of Astronomy, Tsinghua University, Beijing, China}

\icmlcorrespondingauthor{Kangning Diao}{dkn16@foxmail.com}
\icmlcorrespondingauthor{Yi Mao}{ymao@tsinghua.edu.cn}

\icmlkeywords{Machine Learning, ICML}

\vskip 0.3in
]



\printAffiliationsAndNotice{} % otherwise use the standard text.

\begin{abstract}
Large-scale numerical simulations ($\gtrsim 500\rm{Mpc}$) of cosmic reionization are required to match the large survey volume of the upcoming Square Kilometre Array (SKA). We present a multi-fidelity emulation technique for generating large-scale lightcone images of cosmic reionization. We first train generative adversarial networks (GAN) on small-scale simulations and transfer that knowledge to large-scale simulations with hundreds of training images. Our method achieves high accuracy in generating lightcone images, as measured by various statistics with mostly percentage errors. This approach saves computational resources by 90\% compared to conventional training methods. Our technique enables efficient and accurate emulation of large-scale images of the Universe.
\end{abstract}

\section{Introduction}
\label{sec:intro}
In preparation for the upcoming era of 21 cm cosmology, many models have been developed to extract information from observations. These models range from the semi-numerical simulation, e.g. \cmfst\ \citep{Mesinger_2010,Murray_2020} to hydrodynamical radiation transfer simulation, e.g. \ttan \citep{Kannan_2021}, with varying levels of accuracy and computational cost. In addition, different approaches have been applied to infer cosmological and astrophysical parameters, including the Markov Chain Monte Carlo (MCMC) code, e.g. \cmmc\ \citep{Greig_2017} to the machine learning boosted simulation-based inference \citep[e.g.][]{Alsing_2019,Zhao_2022B}. However, parameter inference typically requires many forward simulations. Given the large field of view of the next-generation telescopes, large-scale simulations are required to fully exploit the information contained in the observations. However, these large-scale simulations are computationally expensive, which has inspired the development of emulators as an alternative.

\begin{figure*}[h]
\centering
 \includegraphics[width=0.75\linewidth]{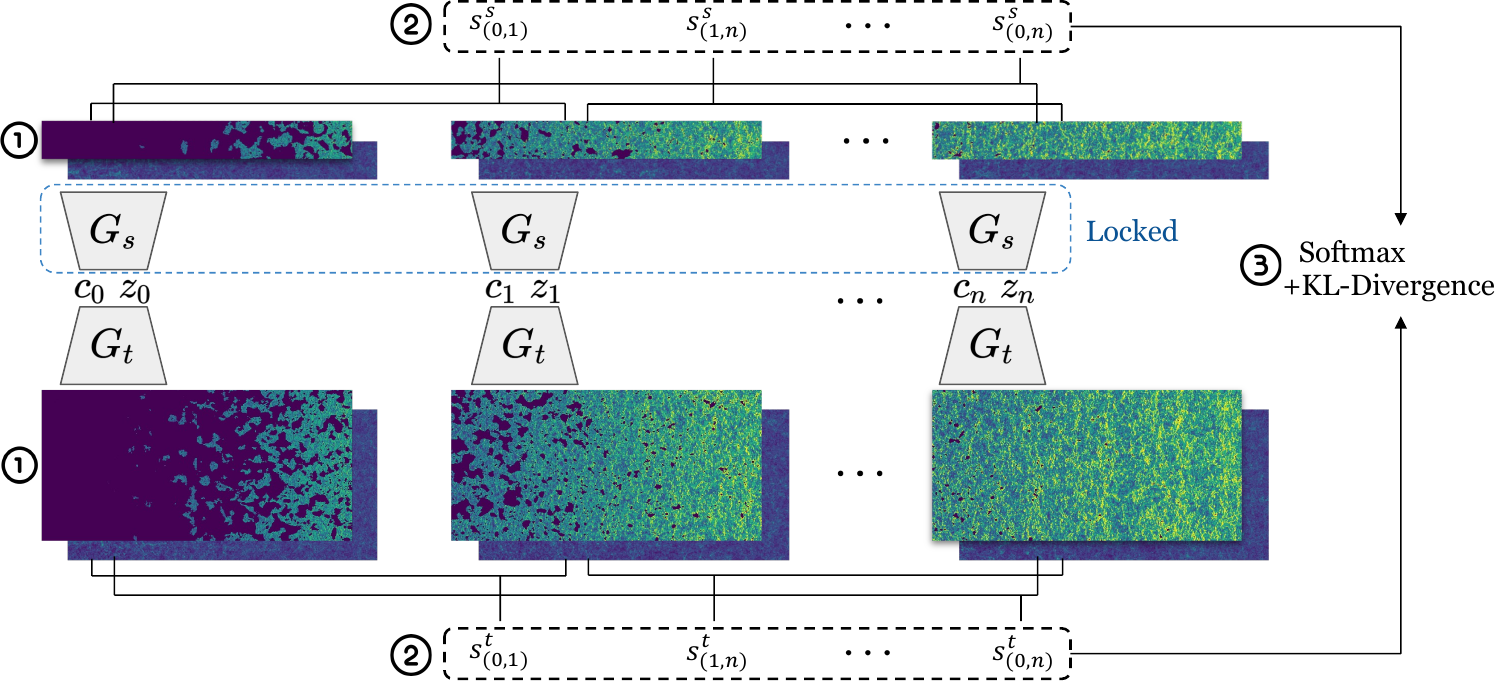}
 \caption{An illustration of the cross-domain correspondence (CDC). \ding{172} We first generate a set of samples with both small-scale GAN and large-scale GAN. \ding{173} We then calculate the similarity between each image pair generated by the same GAN. \ding{174} Finally, we normalize the similarity vector of each GAN with softmax, then compute the KL-divergence as the CDC. The samples shown here are results from our large-scale GAN.}%A summary of my own
 \label{fig:cdc}
\end{figure*}

Building emulators typically requires numerous training samples. For large-scale simulations, the cost of obtaining these training samples can be prohibitive in and of itself. To address this issue, the concept of multi-fidelity emulation \citep{Kennedy_2000,Ho_2021} has been proposed. This approach first uses low-cost (low-fidelity) simulations to create an emulator. The emulator is then calibrated with a small number of high-cost (high-fidelity) simulations, reducing the computational cost while still maintaining the output quality. 

Here we choose GAN \cite{Goodfellow_2014,List_2020,Sambatra_2022} as our emulation model. GAN emulation has previously demonstrated the ability to produce high-quality samples. However, GAN training is known to suffer very often from mode collapse, especially with a dataset smaller than $\sim 1000$ images. In the context of 21 cm lightcone emulation, this would typically require $\gtrsim 1000$ expensive simulations which are sometimes impossibly costly. In this paper, we propose the few-shot transfer learning \citep[e.g.][]{Ojha_2021} to train a faithful large-scale 21 cm lightcone image emulator with a limited number of simulations. Few-shot transfer learning allows us to learn a new task with a limited number of samples, which serves as the `calibrating' procedure in multi-fidelity emulation. This multi-fidelity emulation allows us to significantly reduce the number of simulations required to train an accurate lightcone image emulator.

\section{Methodology}
Our approach involves a two-step process. First, we train our GAN with 120000 small-scale (size of $(2,64,512)$) images. In the second step, we train our large-scale GAN on 320 large-scale (size of $(2,256,512)$) images while preserving the diversity of GAN results. We will explain our approach in detail in the following.

\textbf{StyleGAN 2}:
The GAN architecture used in this work is StyleGAN 2 \citep{Karras_2019}. Our generator $G$ consists of two parts: First, a mapping network $f$ takes the astrophysical parameter $\mathbf{c}$ and a random vector $\mathbf{z}$ and returns a style vector $\mathbf{w}$. Second, a synthesis network $g$ uses the style vector $\mathbf{w}$ to shift the weights in convolution kernels, and Gaussian random noise is injected into the feature map right after each convolution to provide variations in different scales of features. Our discriminator $D$ has a ResNet \cite{He_2015}-like architecture. 
%The most naive adversarial loss is:
%\begin{equation}
%    \mathcal{L}_{\rm adv} = D(G(\mathbf{z},\mathbf{c})|\mathbf{c}) - D(\mathbf{x}|\mathbf{c})\\
%\end{equation}
%
% An $r_1$ loss $\mathcal{L}_{r_1}$ \citep{Lars_2018} is applied to the discriminator to enhance the sparsity of weight matrices, alleviating overfitting. A path-length loss is applied to the generator, which has the form of:
%\begin{equation}
%    \mathcal{L_{\mathrm{path}}} = \left(\Big|\Big|\frac{\partial G(\mathbf{w})}{\partial \mathbf{w}}^{T}G(\mathbf{w})\Big|\Big|_2-a\right)^2
%\end{equation}
%a constant difference $a$ in practice results in more reliable and consistently behaving models \citep{Karras_2019}. Thus, our final training objective is:

%\begin{equation}
%    \begin{aligned}
%        (G^*,D^*) &= \arg \min\limits_{G} \max \limits_{D} \mathbb{E}_{\mathbf{z}\sim p(\mathbf{z}),\mathbf{x}\sim p(\mathbf{x})} \mathcal{L}_{\rm adv}\\
 %   &+\arg \min\limits_{D}\mathbb{E}_{\mathbf{z}\sim p(\mathbf{z}),\mathbf{x}\sim p(\mathbf{x})}\mathcal{L}_{r_1}\\
 %   &+\arg\min\limits_{G}\mathbb{E}_{\mathbf{z}\sim p(\mathbf{z}),\mathbf{x}\sim p(\mathbf{x})}\mathcal{L_{\mathrm{path}}}
 %   \end{aligned}
%\end{equation}

\textbf{Cross-Domain Correspondence (CDC)}: Assuming we have a good small-scale StyleGAN emulator, we expand the size of the generator's first layer, resulting in a final output size of $(2,256,512)$. 

Next, we retrain our GAN with large-scale images. We first employ the patchy-level discriminator and cross-domain correspondence as described in \citet{Ojha_2021}. We mark the small-scale GAN as our source model $G_s$ and the large-scale GAN as the target model $G_t$. First, we use the same batch of vectors $(\mathbf{z},\mathbf{c})$ feeding both $G_s$ and $G_t$, getting the corresponding small-scale images $G_s(\mathbf{z},\mathbf{c})$ and large-scale $G_t(\mathbf{z},\mathbf{c})$. Then we calculate the cosine similarity $s_{(i,j)}$ between any pair of images in $G_s(\mathbf{z},\mathbf{c})$ as
\begin{equation}
    \mathbf{S}_s(\mathbf{z},\mathbf{c})=\{\cos(G_s(z_i,c_i),G_s(z_j,c_j))_{\forall i\neq j}\}
\end{equation}
and similarly for $G_t$ we have:
\begin{equation}
    \mathbf{S}_t(\mathbf{z},\mathbf{c})=\{\cos(G_t(z_i,c_i),G_s(z_j,c_j))_{\forall i\neq j}\}
\end{equation}
Here the $\cos$ denotes the cosine similarity. Next, we normalize these two vectors using softmax and calculate the KL divergence between vectors:
\begin{equation}
    \mathcal{L}_{\rm CDC} = D_{\rm KL}\left(\mathrm{Softmax}(\mathbf{S}_s),\mathrm{Softmax}(\mathbf{S}_t)\right)
\end{equation}
In this way, one can encourage the $G_t$ to generate samples with a diversity similar to $G_s$, relieving the mode collapse problem.

\textbf{Other Techniques}: A patchy-level discriminator is also adopted in this work. We divided the astrophysical parameter space into two parts: the anchor region and the rest. The anchor region is a spherical region around training set parameters with a small radius. In this region, the GAN image $G_t(\mathbf{z},\mathbf{c}_{\rm anch})$ has a good training sample to compare with. Thus, we apply the full discriminator with these parameters. If $\mathbf{c}$ is located outside the anchor region, we only apply a patch discriminator: in this case, the discriminator does not calculate the loss of the whole image but calculates the loss of different patches of the image.

Since the small-scale information in both training sets is identical, we freeze the first two layers of the discriminator \citep{Mo_2020}. We add the small-scale discriminator $D_s$ loss to ensure the correctness of small-scale information. Our code is public-available in this GitHub repo\footnote{\url{https://github.com/dkn16/multi-fidel-gan-21cm}}.

\begin{figure*}
\centering
 \includegraphics[width=0.8\linewidth]{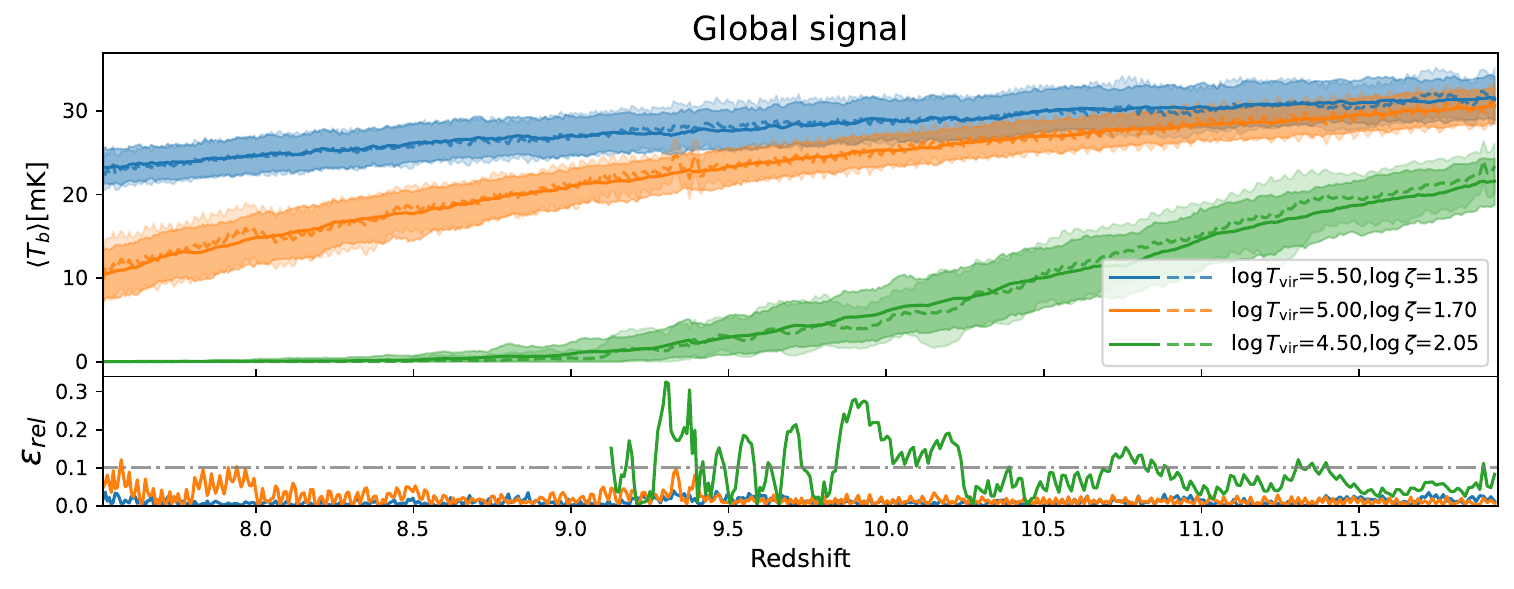}
 \caption{\textit{The upper panel}: global signal reproduced with large-scale GAN. Different colors denote different parameters, the solid line is calculated with the test set, while the dashed line is the GAN result. The shallow shaded region is the $2\sigma$ scatter of the GAN images, while the thick shaded region is the $2\sigma$ scatter of the test set images. \textit{The lower panel}: relative error between GAN global signal and test set global signal, the grey dot-dashed line represents the 10\% error line, while the data points near 0 are neglected.}%A error bar, 3 small samples, sample legend in the left 
 \label{fig:glob256}
\end{figure*}

\begin{figure*}
\begin{center}
%\footnotesize
\begin{tabular}{ccc}
\includegraphics[width=0.3\linewidth]{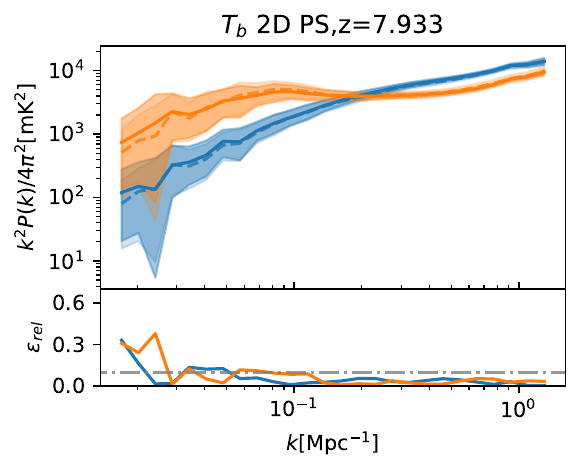}
\includegraphics[width=0.3\linewidth]{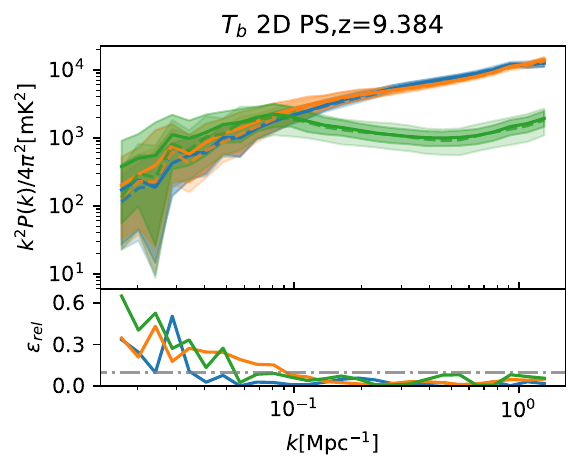}
\includegraphics[width=0.3\linewidth]{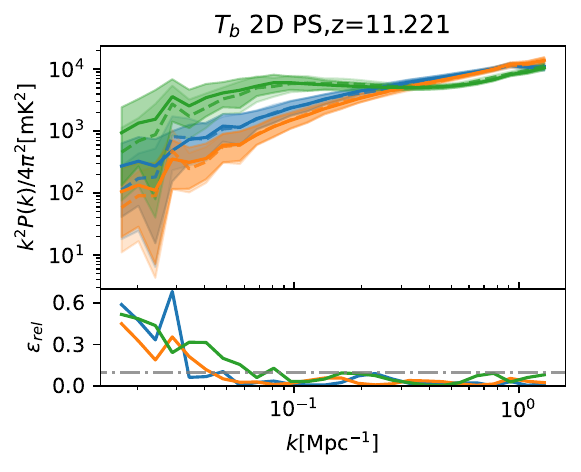}\\
\includegraphics[width=0.3\linewidth]{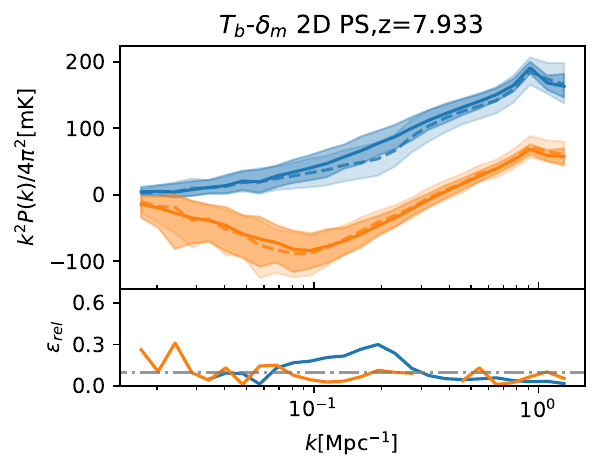}
\includegraphics[width=0.3\linewidth]{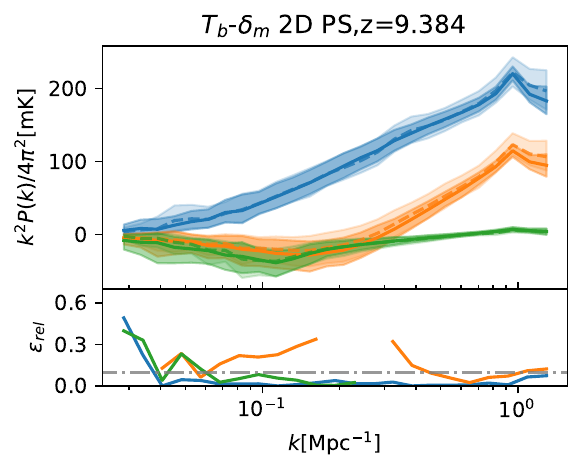}
\includegraphics[width=0.3\linewidth]{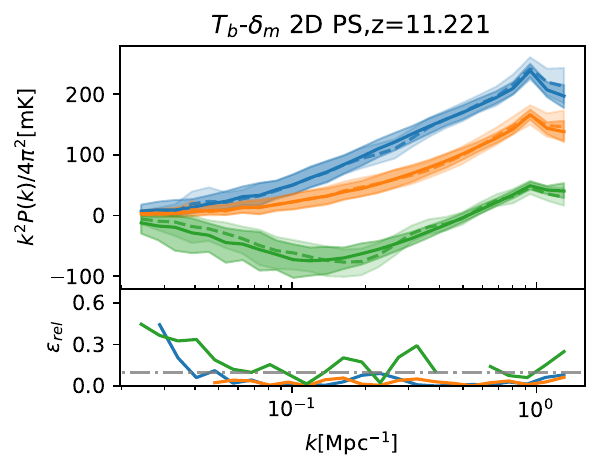}\\
\includegraphics[width=0.3\linewidth]{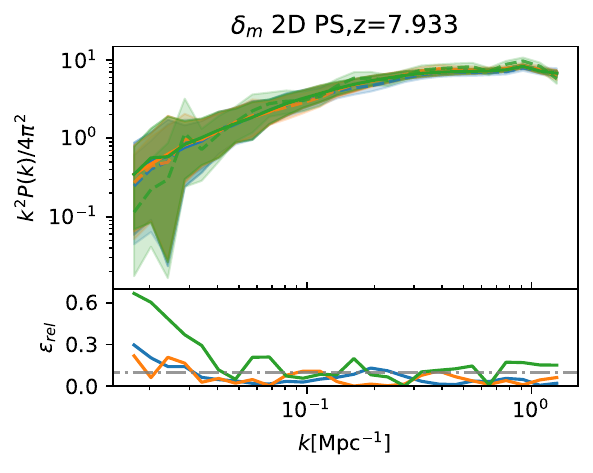}
\includegraphics[width=0.3\linewidth]{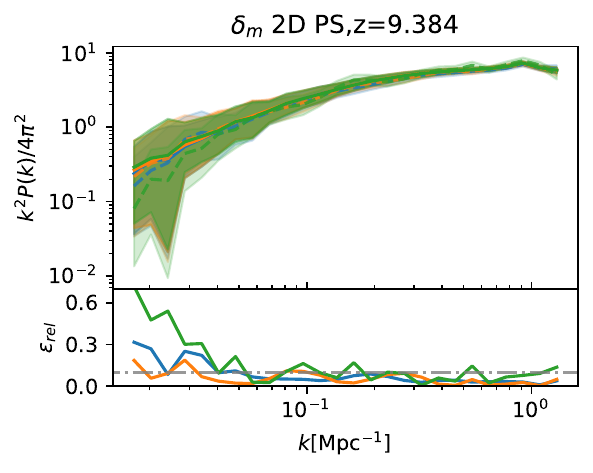}
\includegraphics[width=0.3\linewidth]{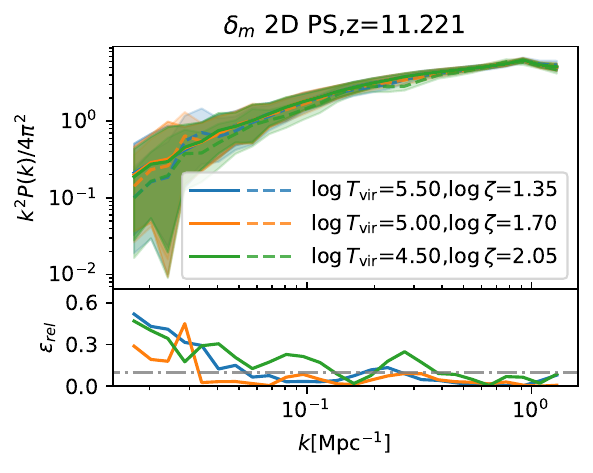}\\
\end{tabular}
\end{center}
\caption{\textit{The upper panel}: the 2D power spectrum of GAN results versus test set PS, each calculated with clips of size (2,256,128). The redshift denotes the redshift of the center slice. Legends are the same as Fig. \ref{fig:glob256}. \textit{The lower panel}: the relative error, same as Fig. \ref{fig:glob256} but for power spectrum here.}% unified error bar. 20% bugget. 
\label{Fig:PS256}
\end{figure*}

\begin{figure}[ht]
\begin{center}
%\footnotesize
\begin{tabular}{c}
\includegraphics[width=0.85\linewidth]{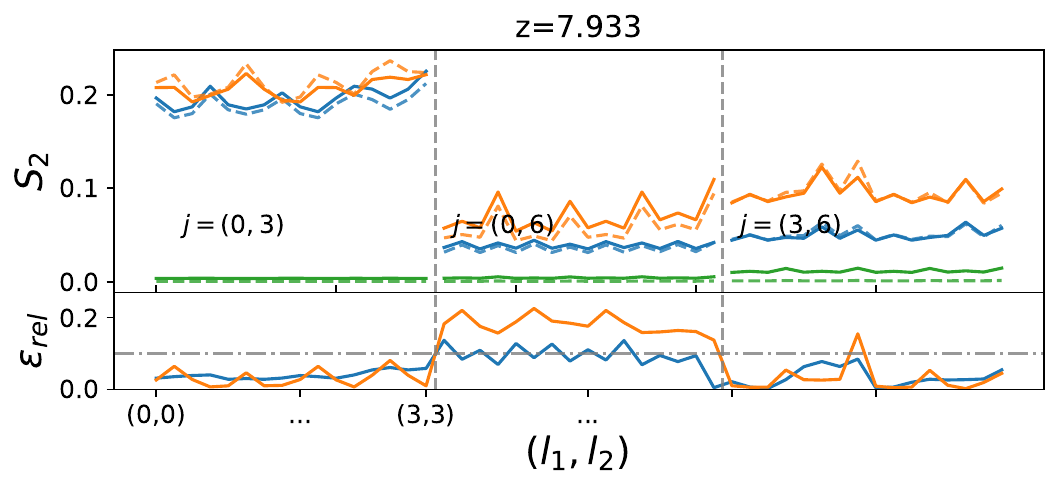}\\
\includegraphics[width=0.85\linewidth]{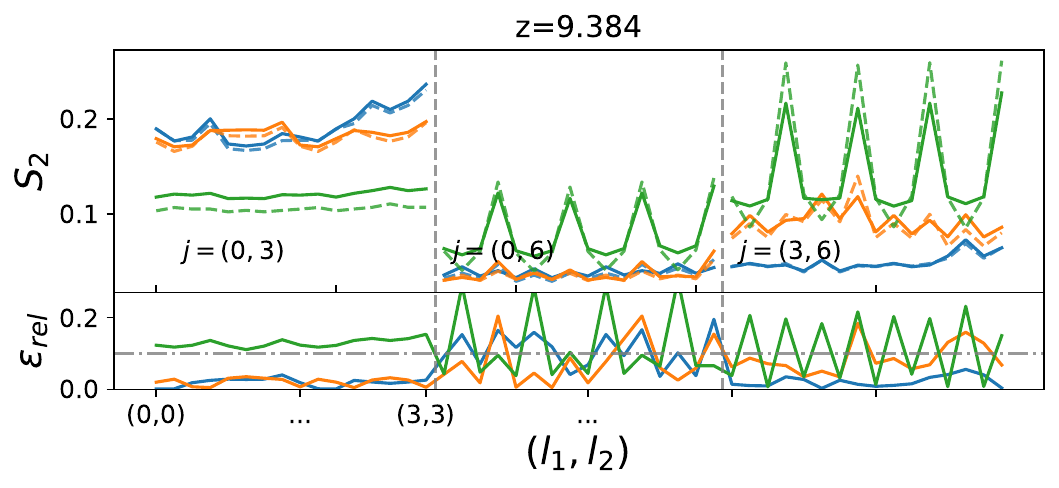}\\
\includegraphics[width=0.85\linewidth]{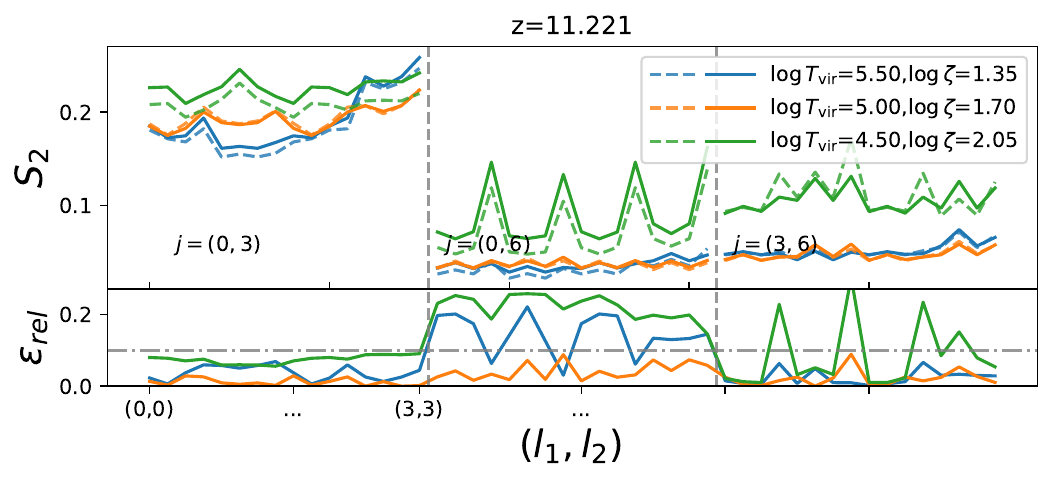}
\end{tabular}
\end{center}
\caption{\textit{The upper panel}: the second order scattering coefficients of GAN results versus test set result, each calculated with clips of size (2,256,128). The redshift denotes the redshift of the center slice. Legends are the same as Fig. \ref{fig:glob256}. \textit{The lower panel}: relative error, same as Fig. \ref{fig:glob256} but for second order scattering coefficients here.}
\label{Fig:S2256}
\end{figure}

\section{Dataset} 
The training dataset for this project consists of two parts: a small-scale dataset and a large-scale dataset. All the data are generated with \cmfst\cite{Mesinger_2010,Murray_2020}, and each simulation has distinct reionization parameters. Our parameters are the ionizing efficiency $\zeta$ and the minimum virial temperature $T_{\rm vir}$. We explored a range of $10<\zeta<250$ and $4<\log T_{ \rm vir}<6$, and the parameters are sampled with Latin-Hypercube Sampling\citep{Mckay_2000}.

The small-scale dataset has a resolution of $(64,64,512)$ and consists of 30,000 simulations with a comoving box length of $(128,128,1024)\rm Mpc$. The third axis ($z$-axis) is along the line of sight (LoS), spanning a redshift range of $7.51<z<11.93$. For each redshift, we run a realization
and select the corresponding slice for our final data. We include the matter overdensity field $\delta_m$ and the 21 cm brightness temperature $T_b$ field for training. Since the overdensity field is highly correlated with other intensity mappings (IM) like CO and C[II] lines, we expect our method can be transferred to other IM images smoothly. For each sample, we cut four image slices, resulting in 120000 lightcone images with a size of $(2,64,512)$ in our small-scale dataset, containing both the overdensity and brightness temperature field.

The large-scale dataset has a $(256,256,512)$ resolution and consists of 80 simulations with a comoving box length of $(512,512,1024)\rm Mpc$, covering the same redshift range. As before, for each sample, we cut four slices and obtained 320 lightcone images with a size of $(2,256,512)$ in our large-scale dataset.

%Our parameters are the ionizing efficiency $\zeta$ and the minimum virial temperature $T_{\rm vir}$. We explored a range of $10<\zeta<250$ and $4<\log T_{ \rm vir}<6$.

\section{Results}
\label{sec:res}
Here we present the evaluation of our model results. A visual inspection of generated samples is shown in Fig. \ref{fig:cdc}. We tested our result on 3 combinations of parameters, each having distinct evolution history. For each parameter combination, we run 4 simulations with distinct initial conditions generated with different random seeds for testing.

\textbf{Global Signal}:
We calculated the global 21 cm signal of the GAN results. Limited by the size of the test set, the mean value is calculated with 1024 image samples. Our result is shown in Fig. \ref{fig:glob256}. We see that GAN works well, with an error of mostly less than 5\% and a well-matched $2\sigma$ region. 

\textbf{Power spectrum (PS)}: 
Fig. \ref{Fig:PS256} shows the $T_b$ auto-PS, $T_b-\delta_m$ cross-PS and $\delta_m$ auto-PS. GAN results perform well on small scales, with an error of less than $10\%$, except when the PS is close to 0. On extremely large scales, the error can exceed $50\%$. This is unsurprising because we lack training samples. The GAN still captures the large-scale power when the $T_b$ signal has a high amplitude. Moreover, the relative error is insignificant compared with the sampling variance.

From $T_b$ auto-PS figures (Fig. \ref{Fig:PS256}, top row), the change of lines shows an evolution with the time that power is transferred from small scale to large scale. Again, the accuracy of the cross-PS (Fig. \ref{Fig:PS256}, middle row) guarantees the correlation between $T_b$ and $\delta_m$. At early stages, the HI traces the matter field well, and the GAN $T_b$ and $\delta_m$ fields have positive cross-correlation at all scales. Later, the cross-correlation becomes negative due to the fact that dense regions hosted ionizing sources earlier and ionized first. Our GAN performs well in reproducing these features. The GAN samples with different parameters have similar matter PS (Fig. \ref{Fig:PS256}, bottom row), which agrees with the truth.

\textbf{Non-Gaussianity}: 
Here we employ the scattering transform \citep[ST, e.g.][]{Mallat_2012,Allys_2019,Cheng_2020,Greig_2022} coefficients as a non-Gaussian statistic to evaluate our GAN. A detailed description can be found in e.g. \citet{Cheng_2021}. We calculated the second-order ST coefficients $S_2$ as measures for non-Gaussianity with \textsc{Kymatio} \citep{Andreux_2020}. As the image sample size grows, we set the kernel size scale $j=0,3,6$ to capture more large-scale information. Results are shown in Fig. \ref{Fig:S2256}. When $(j_1,j_2) = (0,3)$, the error is less significant as $\lesssim 10\%$. When $j_2=6$ the error exceeds $20\%$.

\section{Summary}
In this paper, we introduce the few-shot transfer learning technique to build an emulator for large-scale 21 cm simulations. The large-scale GAN is trained with 80 simulations, and the relative error of statistics is less than $10\%$ on small scales. On large scales, a mild increase in error arises due to insufficient training samples.

Generating our multi-fidelity dataset requires $\sim 1.2\times 10^5$ CPU hours, while  a purely large scale dataset requires $\sim 1.5\times 10^5$ CPU hours, with 5000 simulations, an optimistic estimate of dataset size consistent with e.g. \citet{Hassan_2022,Sambatra_2022}. Our method reduces the computational cost by 90\%, which will enable us to emulate more complex simulations in the future.

\section*{Acknowledgements}

This work is supported by the National SKA Program of China (grant No. 2020SKA0110401), NSFC (grant No. 11821303), and the National Key R\&D Program of China (grant No. 2018YFA0404502). We thank Xiaosheng Zhao, Ce Sui, and especially Richard Grumitt for inspiring discussions. 
We acknowledge the Tsinghua Astrophysics High-Performance Computing platform at Tsinghua University for providing computational and data storage resources that have contributed to the research results reported within this paper. 

\bibliography{example_paper,reference}
\bibliographystyle{icml2023}

%%%%%%%%%%%%%%%%%%%%%%%%%%%%%%%%%%%%%%%%%%%%%%%%%%%%%%%%%%%%%%%%%%%%%%%%%%%%%%%
%%%%%%%%%%%%%%%%%%%%%%%%%%%%%%%%%%%%%%%%%%%%%%%%%%%%%%%%%%%%%%%%%%%%%%%%%%%%%%%
% APPENDIX
%%%%%%%%%%%%%%%%%%%%%%%%%%%%%%%%%%%%%%%%%%%%%%%%%%%%%%%%%%%%%%%%%%%%%%%%%%%%%%%
%%%%%%%%%%%%%%%%%%%%%%%%%%%%%%%%%%%%%%%%%%%%%%%%%%%%%%%%%%%%%%%%%%%%%%%%%%%%%%%
\newpage
\appendix
%\onecolumn
\section{Comparison with previous work}
\begin{table*}

 \caption{A comparison of accuracy and cost of various methods. Large-scale GAN is training GAN with only large-scale image samples, and the performance is estimated. The cost here is the CPU hours used to generate the dataset.}
 \label{tab:bench}
 \centering
 \begin{minipage}{\linewidth}
 \begin{tabular*}{\linewidth}{lcccr}
  \hline
  Method & \multicolumn{2}{c}{Relative Error}& Cosmic Variance & Cost [CPU hours]\\
   & At small scales & At large scales & \\
  \hline
  Small-scale GAN & $<10\%$ & --- & Large&$1\times 10^4$\\
  Large-scale GAN (estimated\footnote{According to related work, 5000 large-scale simulations is an optimistic estimation for  the necessary training samples \citep[e.g.][]{Hassan_2022,Sambatra_2022}, which will take $1.5 \times 10^5$ CPU hours; to ensure the $10\%$ accuracy and to make  a fair comparison, 30000 simulations are required, which will cost $9\times 10^5$ CPU hours.}) &$<10\%$& $<10\%$& Small & $(1.5 - 9)\times 10^5$\\
 Few-shot GAN (this work) &$<10\%$ & $20\%-50\%$ & Small & $1.2\times 10^4$\\
  \hline
 \end{tabular*}
 \end{minipage}
\end{table*}

Several noteworthy applications of GAN in astronomy have been extensively explored in previous studies \citep[e.g.,][]{List_2020, Sambatra_2022, Yiu2022, Tilman2019}. Previous works have made significant progress in utilizing innovative GAN structures such as the progressively growing GAN (PGGAN) \citep{Karras2017} and stabilized GAN \citep{Liu2021}. These studies have demonstrated sub-percent-level accuracy, as assessed by various statistical measures, for unconditional emulation, and achieved accuracy at the ten percent level for conditional emulation. A comparison between our results and previous findings is presented in Table \ref{tab:bench}. By employing the StyleGAN2 architecture, we have achieved percent-level accuracy in conditional emulation with sufficient training samples, as validated by various statistical measures. In the few-shot learning scenario, our GAN exhibits similar accuracy on a small scale and demonstrates a moderate increase on a larger scale. Furthermore, our large-scale GAN, combined with few-shot transfer learning techniques, allows for computational resource savings ranging from 90\% to 99\%, depending on different estimations.

\section{Test on mode collapse}
\subsection{Visual inspection}
\begin{figure*}[h]
\centering
 \includegraphics[width=0.95\linewidth]{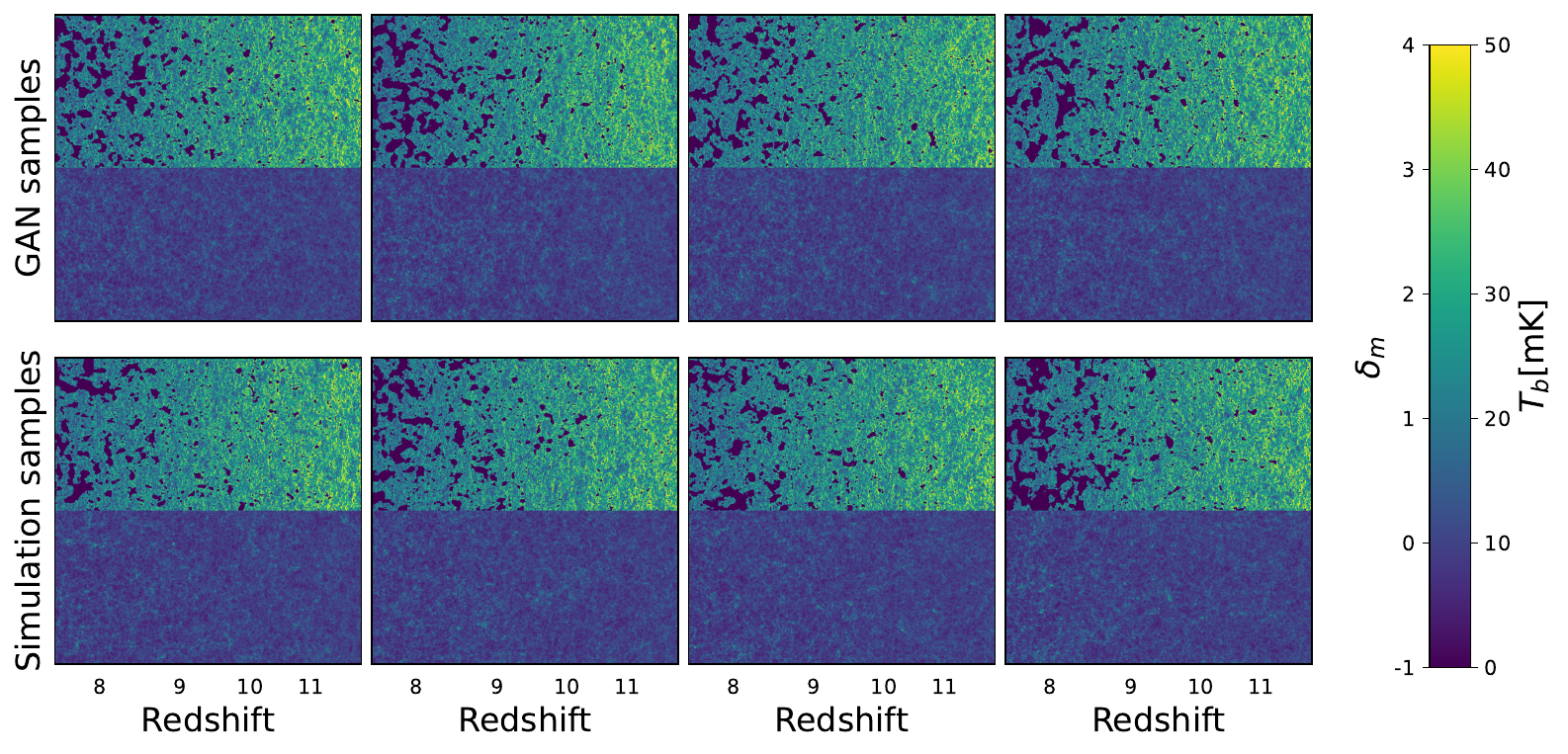}
 \caption{Here we present four distinct realizations for both GAN and simulation samples, allowing for visual inspection. These realizations were generated with the parameters $\log T_{\rm vir} = 5.50$ and $\log \zeta = 1.70$. The upper panel displays the GAN samples, while the lower panel showcases the simulation samples. Each realization was computed using a unique latent vector (for GAN) or initial condition (for simulation). Within each subplot, the upper half represents the brightness temperature field, while the lower half represents the overdensity field.}%A summary of my own
 \label{fig:vis}
\end{figure*}

To assess the diversity of our model, we conducted a visual inspection. We generated multiple realizations for both GAN samples and simulation samples, as illustrated in Fig. \ref{fig:vis}. Upon careful observation, we observed that the shape and size of ionized bubbles exhibit variation across different GAN samples, indicating the absence of any specific preference for bubble features. Furthermore, the locations of ionized bubbles also appear random, as no discernible trend or pattern was observed among the samples we examined.

\subsection{Pixel level variance}
\begin{figure*}[h]
\centering
 \includegraphics[width=0.95\linewidth]{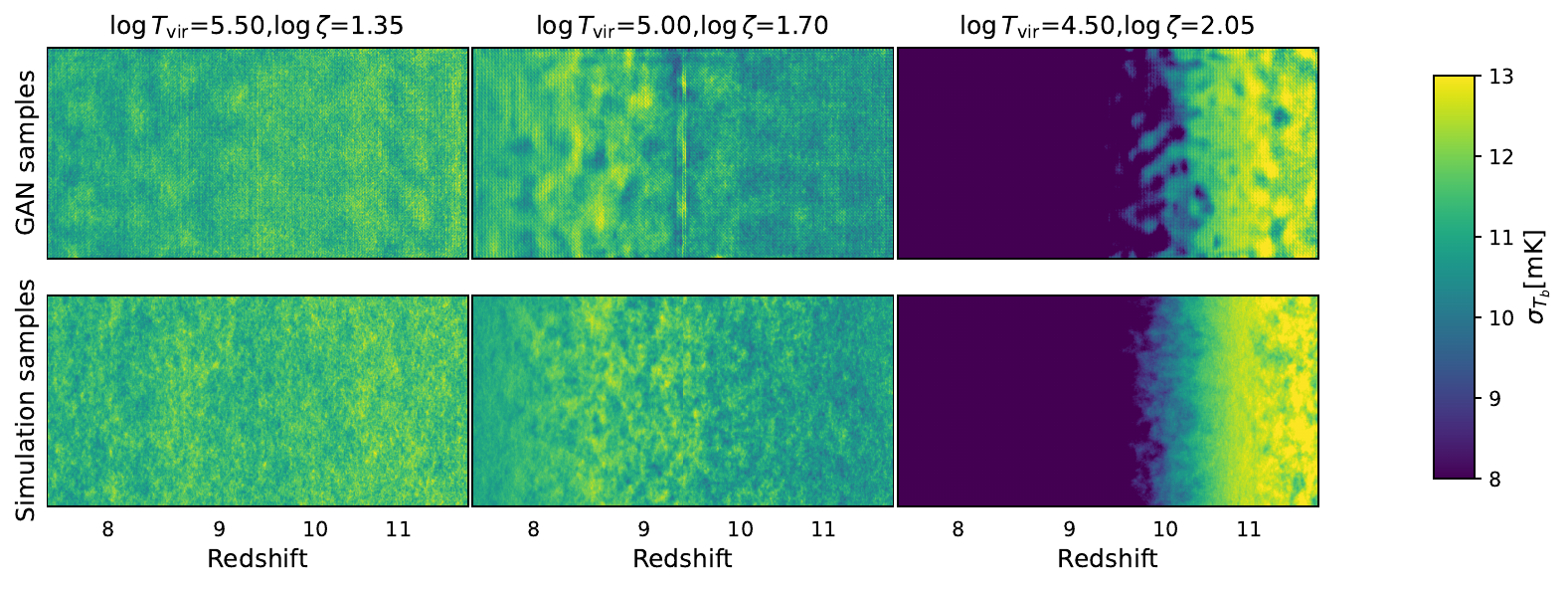}
 \caption{The standard deviation in each pixel is calculated for the 21 cm $T_b$ map. The upper panel is for the GAN samples while the lower panel is for the simulation samples.}%A summary of my own
 \label{fig:var}
\end{figure*}
In addition to visual inspections, we also computed the standard deviation of the $T_b$ field for each pixel, as depicted in Fig. \ref{fig:var}. Our aim was to observe any potential decrease in the standard deviation, which could indicate mode collapse. Upon analyzing the results in Fig. \ref{fig:var}, we noticed that the variance for both GAN and simulation samples appeared similar, particularly for higher $T_b$ values. However, we observed mild fluctuations in the standard deviation when the $T_b$ value was low. Based on this analysis, we can conclude that there is no clear evidence of significant mode collapse at the pixel level.

\subsection{Feature level variance}
\begin{figure}[h]
\begin{center}
%\footnotesize
\includegraphics[width=0.95\linewidth] {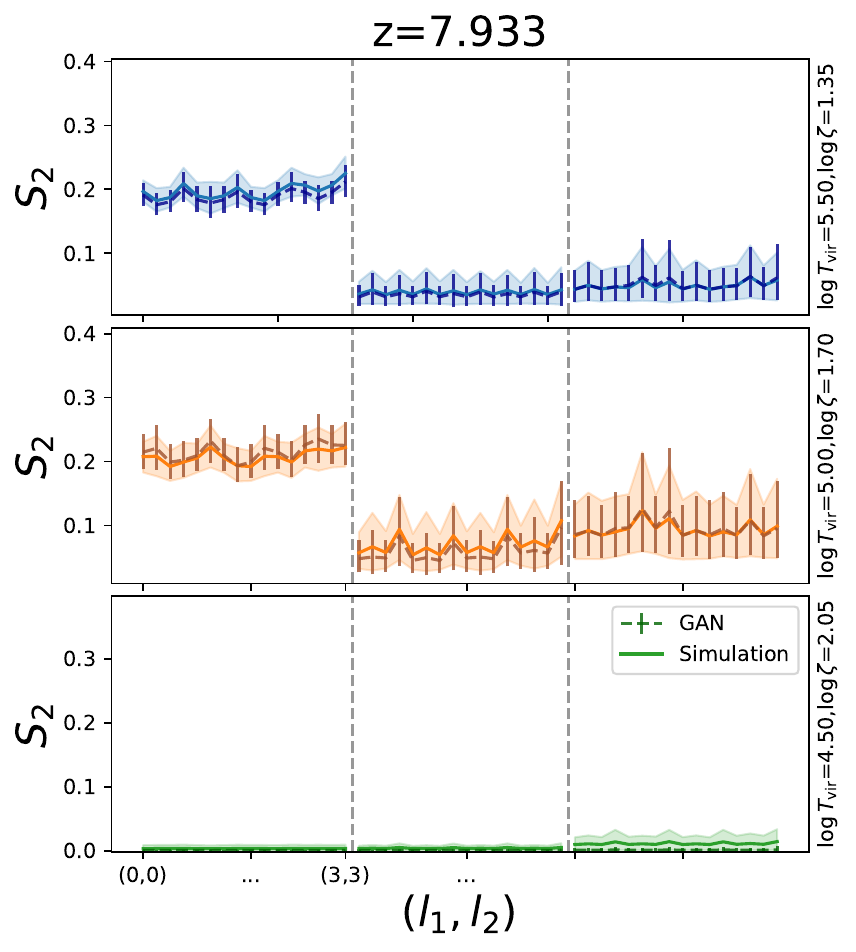}
\end{center}
\caption{The 2$\sigma$ scatter of $S_2$ for in simulations and GAN samples at $z=7.933$. The solid line is for the simulation mean value, while the dashed line is for GAN mean. Shaded region is the $2\sigma$ scatter for simulation samples while the error bar is for GAN samples. Different plot corresponds to different reionization parameters, as is shown on the right of each plot.}
\label{Fig:S2std2561}
\end{figure}
\begin{figure}[h]
\begin{center}
%\footnotesize
\includegraphics[width=0.95\linewidth] {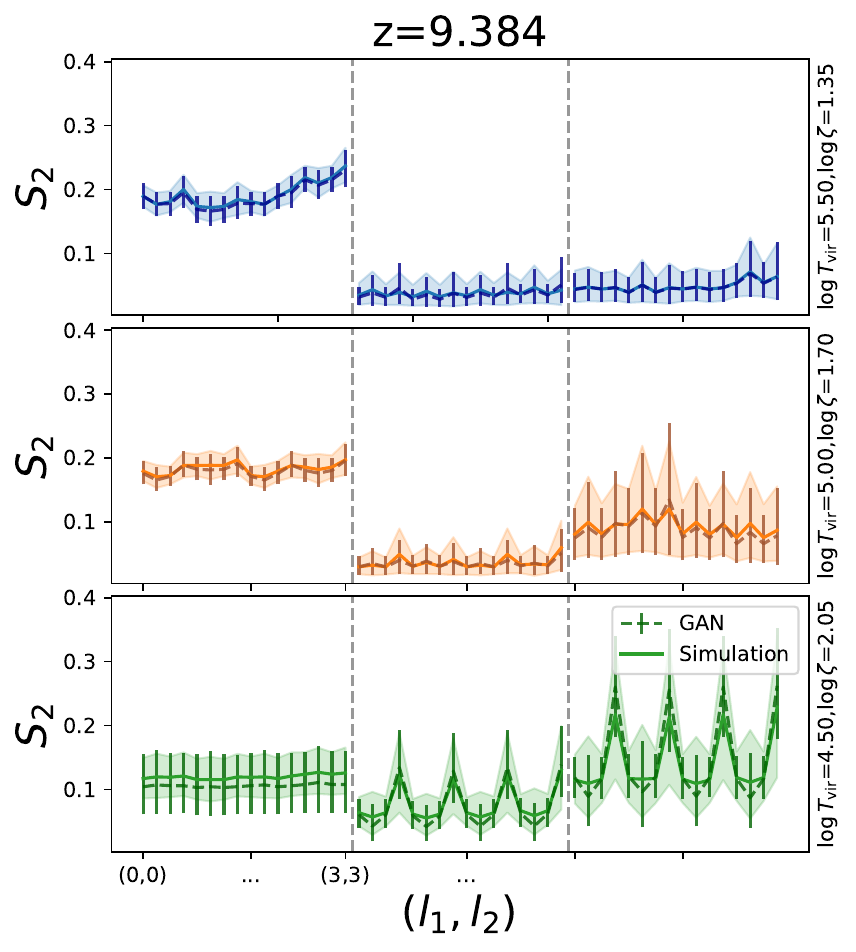}
\end{center}
\caption{Same as Fig. \ref{Fig:S2std2561}, but for $z=9.384$.}
\label{Fig:S2std2562}
\end{figure}
\begin{figure}[h]
\begin{center}
%\footnotesize
\includegraphics[width=0.95\linewidth] {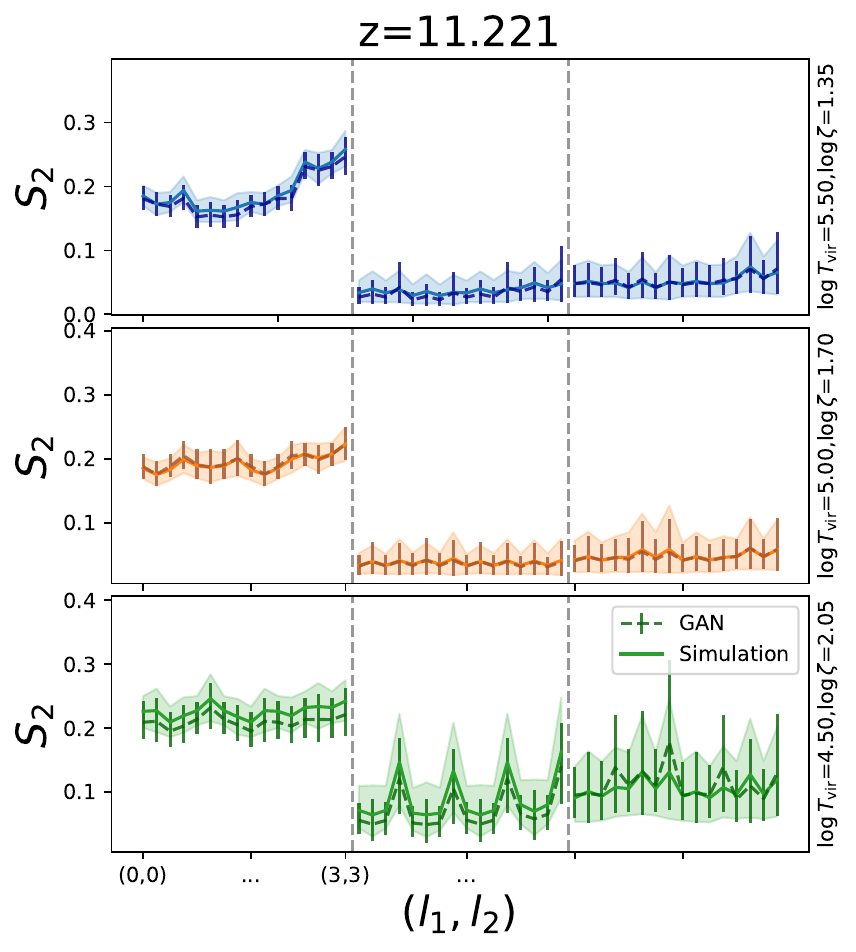}
\end{center}
\caption{Same as Fig. \ref{Fig:S2std2561}, but for $z=11.221$.}
\label{Fig:S2std2563}
\end{figure}
Lastly, we computed the 2$\sigma$ scatter of the second-order ST coefficients ($S_2$) for the $T_b$ field, which serves as a representation of image features. The results are presented in Figures \ref{Fig:S2std2561}-\ref{Fig:S2std2563}. Consistent with the analysis in Section \ref{sec:res}, we selected the scales $(j_1,j_2)$ as (0,3), (0,6), and (3,6) to capture both small and large-scale features.

Upon examination, we observed that in most cases, the $2\sigma$ scatter of GAN features overlapped with that of simulation samples, indicating the absence of mode collapse at the feature level. However, in the bottom subplot of Fig. \ref{Fig:S2std2562}, we noticed a deviation in both the mean value and $2\sigma$ scatter for certain features at the super-large scale. This suggests a slight mode collapse issue in the generated images at that particular scale.

In conclusion, our analysis indicates that there is no strong evidence of mode collapse at the feature level. The GAN samples generally mimic the behavior of the simulation samples quite well, except when the $T_b$ approaches zero.
%%%%%%%%%%%%%%%%%%%%%%%%%%%%%%%%%%%%%%%%%%%%%%%%%%%%%%%%%%%%%%%%%%%%%%%%%%%%%%%
%%%%%%%%%%%%%%%%%%%%%%%%%%%%%%%%%%%%%%%%%%%%%%%%%%%%%%%%%%%%%%%%%%%%%%%%%%%%%%%

\end{document}